\documentclass[aps,pre,twocolumn,superscriptaddress]{revtex4}

\usepackage{color}
\usepackage{amsmath}      
\usepackage{amssymb} 
\usepackage{amsfonts}   
\usepackage{graphicx}
\usepackage{pgfplots}
\usepackage{bm}
\usepackage{ifthen}
\newboolean{pnas}
\setboolean{pnas}{false}

\usepackage{bbm}
\usepackage{textcomp}
  
\graphicspath{{../Fig1_Framework/}{../Fig2_Setup/}{../Fig3_Inference/}{../Fig5_Decoding/}{../Fig4_Sensitivity/}{../Fig6_Efficient/}{../Fig_SI_trajectories/}{../Fig_SI_simulations/}{../Fig_SI_summaryRef2/}}

\newcommand{\beq}{\begin{eqnarray}}
\newcommand{\eeq}{\end{eqnarray}}
\newcommand{\<}{\langle}
\renewcommand{\>}{\rangle} 

\newcommand{\s}{\sigma}

\newcommand{\myparagraph}[1]{\noindent\paragraph*{#1}}

\begin{document}

\title{Closed-loop estimation of retinal network sensitivity\\
reveals signature of efficient coding}
\author{Ulisse Ferrari}
\thanks{These authors contributed equally.}
\affiliation{Institut de la Vision, INSERM and UMPC, 17 rue Moreau, 75012 Paris, France}
\author{Christophe Gardella}
\thanks{These authors contributed equally.}
\affiliation{Laboratoire de physique statistique, CNRS, UPMC and \'Ecole normale sup\'erieure, 24, rue Lhomond, 75005 Paris, France}
\affiliation{Institut de la Vision, INSERM and UMPC, 17 rue Moreau, 75012 Paris, France}
\author{Olivier Marre}
\thanks{These authors contributed equally. Correspondence should be sent to \url{olivier.marre@gmail.com} and \url{tmora@lps.ens.fr}.}
\affiliation{Institut de la Vision, INSERM and UMPC, 17 rue Moreau, 75012 Paris, France}
\author{Thierry Mora}
\thanks{These authors contributed equally. Correspondence should be sent to \url{olivier.marre@gmail.com} and \url{tmora@lps.ens.fr}.}
\affiliation{Laboratoire de physique statistique, CNRS, UPMC and \'Ecole normale sup\'erieure, 24, rue Lhomond, 75005 Paris, France}

\begin{abstract}
According to the theory of efficient coding, sensory systems are adapted to represent
natural scenes with high fidelity and at minimal metabolic cost.
Testing this hypothesis for sensory structures performing non-linear computations on high dimensional stimuli is still an open challenge. 
Here we develop a method to characterize the sensitivity of the retinal network to perturbations of a stimulus. Using closed-loop experiments, we explore selectively the space of possible perturbations around a given stimulus. We then show that the response of the retinal population to these small perturbations can be described by a local linear model. 
Using this model, we computed 
the sensitivity of the neural response to arbitrary temporal
perturbations of the stimulus, and found a peak in the sensitivity as
a function of the frequency of the perturbations. Based on a minimal
theory of sensory processing, we argue that this peak is set to
maximize information transmission. Our approach is relevant to 
testing the efficient coding hypothesis locally in any context where
no reliable encoding model is known.

\end{abstract}

\maketitle

\bigskip

The efficient coding hypothesis \cite{Attneave1954,Barlow1961} postulates that neural encoding of stimuli has adapted to represent natural occuring sensory scenes optimally in the presence of limited resources. 
This principle can be recast in the language of information theory, and it has been argued that early stages of sensory processing aim to represent information in an efficient form \cite{Bialek1987,Atick1992,Simoncelli2001}. 

How can we test if this hypothesis is valid? When sensory systems can be approximated as linear or quasi-linear filters \cite{Pitkow2012,Doi2012,Karklin2011}, the efficient coding hypothesis yields quantitative predictions about the shape of the receptive fields \cite{Deco1997,Bell1997,Atick1990,Dan1996,Olshausen1996}. Beyond the linear case, predictions have only been drawn in restricted cases, e.g. for one-dimension stimuli or single neurons \cite{Laughlin1981,Steveninck1997,Brenner2000,Fairhall2001,Schwartz2001,Gutnisky2008,Karklin2009}. 
Testing the efficient coding hypothesis for non-linear systems is more challenging, especially if the stimulus and response are high dimensional. 
If a non-linear model existed that could accurately predict the network response to complex stimuli, then the sensitivity of the representation---its ability to discriminate responses to small perturbations around any given stimulus---could be estimated directly from that model.
According to efficient coding theory, the form of this sensitivity should be related to the stimulus statistics \cite{Brunel1998,Wei2016}. 
However, in the majority of cases, such a comprehensive non-linear model is not available, and there is no method to quantify sensitivity and to assess the efficient coding hypothesis.
Even in the retina, previous studies have shown that the response to complex stimuli can be produced by complex circuits involving several non-linearities \cite{Gollisch2010,Mcintosh2016}, 
and a general predictive model is lacking. 
New ways to test the efficiency of the retinal representation of complex scenes are thus needed. 

Here we present a novel approach to measure experimentally the sensitivity of a non-linear network, and compare it to the prediction derived from the efficient coding hypothesis. Because any non-linear function can be linearized around a given point, we hypothesized that, even in a sensory network with non-linear responses, one can still define a local linear model that can well describe the network response to small perturbations around a given reference stimulus. This local model should only be valid around the reference stimulus, but it is sufficient to estimate the sensitivity of the network. 

We applied this strategy to the retina. We recorded the activity of a large population of retinal ganglion cells stimulated by a randomly moving bar.
We characterized the sensitivity of the retinal population to small stimulus changes, by testing perturbations around a reference stimulus, using closed-loop experiments. This allowed us to build a complete model of the population response in that region of the stimulus space, and to precisely quantify the sensitivity of the neural representation. We found that the sensitivity exhibits a peak as a function of frequency, in agreement with the prediction from efficient coding theory. Our approach is general and can be used for any sensory network, and allows for testing the efficient coding theory even for high dimension stimuli and non-linear networks. 

\bigskip

\section*{Results}
\noindent\myparagraph{\bf Measuring sensitivity using closed-loop experiments.}
We recorded from a population of 60 ganglion cells in the rat retina using a 252-electrode array while presenting a randomly moving bar (see Fig.~1A {and Materials and Methods}). 
{Our aim was to measure the sensitivity of the retinal population to perturbations of a pre-defined stimulus. We measured the response to many repetitions of a short (0.9 s) reference stimulus, as well as many small perturbations around it.}
The reference stimulus was the random trajectory of a white bar on a dark background undergoing Brownian motion with a restoring force (see Materials and Methods).
Perturbations were small changes affecting that reference trajectory in its middle portion, between 300 and 630 ms.
Perturbations varied both in shape and in amplitude: we used 16 different perturbation shapes (shown in Fig.~S1), each presented at different amplitudes ({Fig.~1A}). 
The population response was defined as sequences of spikes and silences in 20 ms {time bins} for each neuron.

\begin{figure}
\centering
\includegraphics[width=1\linewidth]{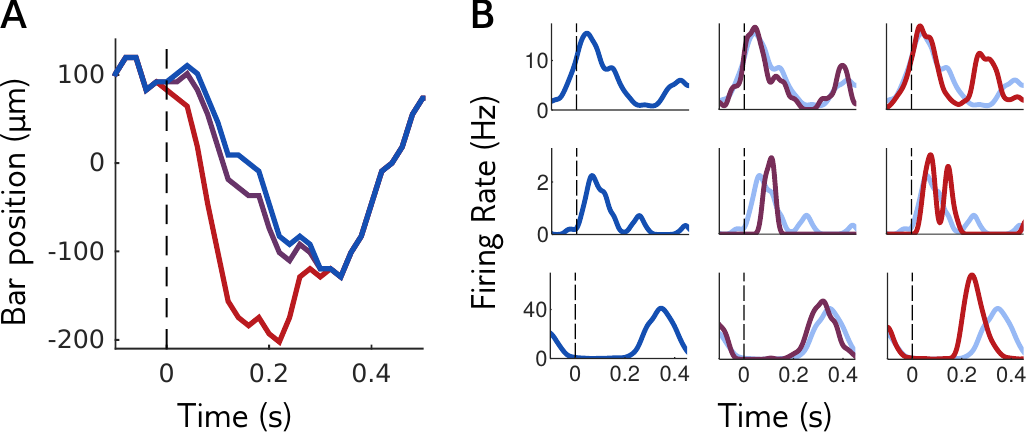}
\caption{
{\bf Sensitivity of a neural population to visual stimuli.}
{\bf A.}: the retina is stimulated with repetitions of a reference
stimulus (here the trajectory of a bar, in blue), and with perturbations of this reference stimulus of different shapes and amplitudes. Purple and red trajectories are perturbations with the same shape, of small and large amplitude. 
{\bf B.}: mean response of three example cells to the reference stimulus (left column and light blue in middle and right columns) and to perturbations of small and large amplitudes (middle and right columns).} 
\label{f:framework}
\end{figure}

To assess the sensitivity of the neural code,
we asked how well different perturbations could be discriminated from the reference stimulus based on the population response.
We expect the ability to discriminate perturbations to depend on their amplitude: responses to small perturbations should be hardly distinguishable, while large perturbations should elicit easily detectable changes, as can be seen in Fig.~1B.
One thus needs to find the range of amplitudes for which discrimination is hard but not impossible. {This requires looking for the adequate range of perturbation amplitudes ``online,'' during the time course of the experiment. }

\begin{figure}
\centering
\includegraphics[width=.9\linewidth]{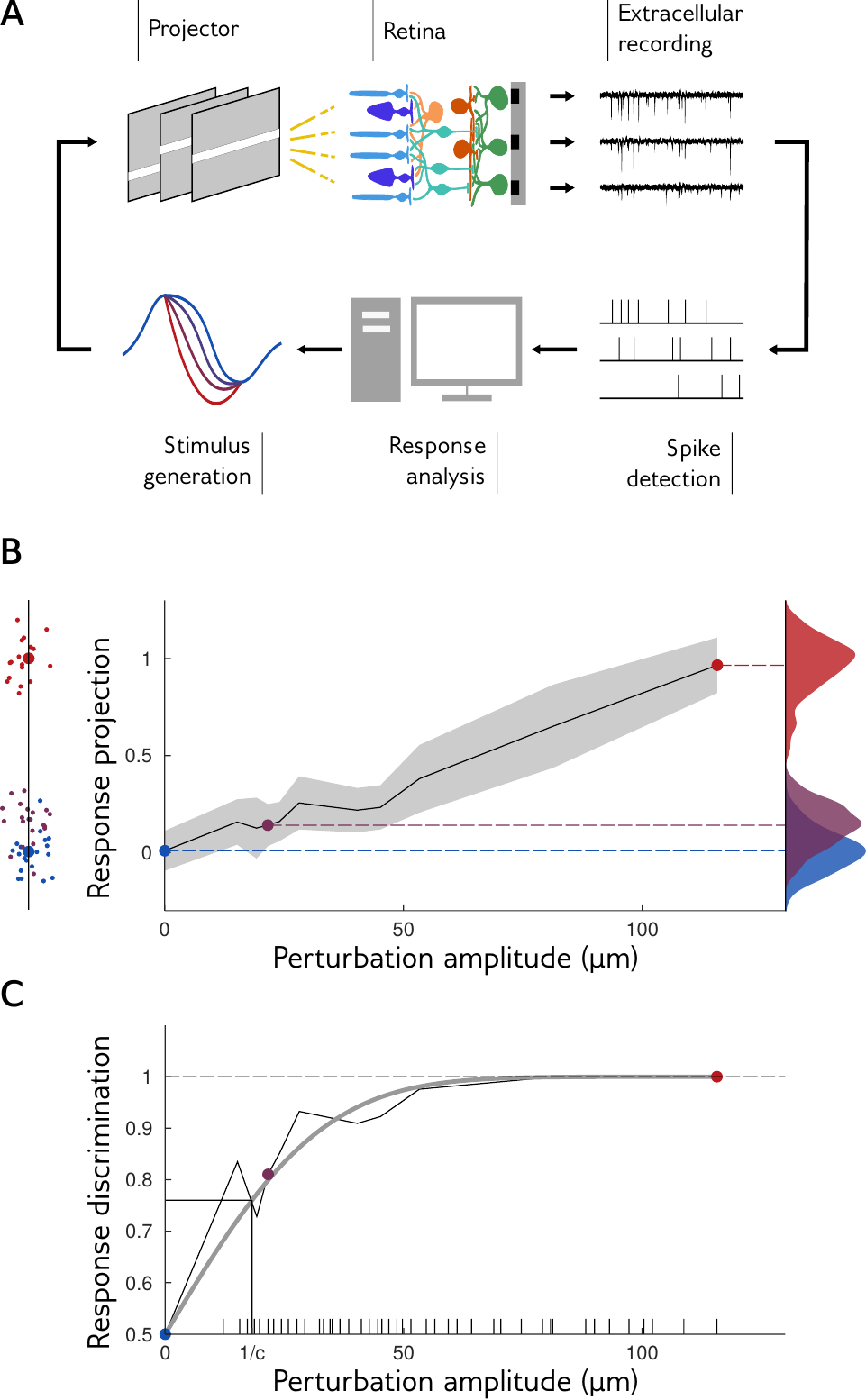}
\caption{ 
{\bf Closed-loop experiments to probe the range of stimulus sensitivity.}
{\bf A.} Experimental setup:  we stimulated a rat retina with a moving bar. Retinal ganglion cell (RGC) population responses were recorded extracellularly with a multi-electrode array. Electrode signals were high-pass filtered and spikes were detected by threshold crossing. We computed the discrimination probability of the population response, and  adapted the amplitude of the next perturbation.
{\bf B.} Left: the neural responses of 60 sorted RCGs are projected along the axis going through the mean response to reference stimulus and the mean response to a large perturbation. Small dots are individual responses, large dots are means. 
Middle: mean and standard deviation (in grey) of response projections for different amplitudes of an example perturbation shape. 
Right: distributions of the projected responses to the reference (blue), and to small (purple) and large (red) perturbations.
Discrimination is high when the distribution of the perturbation is well separated from the distribution of the reference.
{\bf C.} Discrimination probability as a function of amplitude $A$. The discrimination increases as an error function, $(1/2)[1+\mathrm{erf}(d'/2)]$, with $d'=c\times A$ (grey line: fit). Ticks on the x axis show the amplitudes that have been tested during the closed-loop experiment.
}
\label{f:setup}
\end{figure}

In order to adapt the amplitude of perturbations to the sensitivity of responses for each perturbation shape, we implemented closed-loop experiments (Fig.~2A). At each step, the retina was stimulated with a perturbed stimulus and the population response was recorded.
Spikes were detected in real time for each electrode independently by threshold crossing (see Materials and Methods). This coarse characterization of the response is no substitute for spike sorting, 
but it is fast enough to be implemented in real time between two stimulus presentations, and sufficient to detect changes in the response. This method was used to adaptively select the range of perturbations in real time during the experiment. Proper spike sorting was performed after the experiment using the procedure described in \cite{Marre2012,Yger16}, and used for all subsequent analyses.

To test whether a perturbation was detectable from the retinal response, we projected the population responses along {an axis defined by} the difference between the mean response to a large-amplitude perturbation and the mean response to the reference (Fig.~2B). By definition, on average the projected response to a perturbation is larger than to the reference. However, both are noisy and broadly distributed around their mean (see Fig.~2B, right, for example distributions).
We define the discrimination probability as the probability that the response to the perturbation is in fact larger than to the reference. Its value is $100\%$ if the responses to the reference and perturbation are perfectly separable, and $50\%$ if their distributions are identical, in which case the classifier does no better than chance. This discrimination probability is equal to the `area under the curve of the receiver-operating characteristics,' which widely used for measuring the performance of binary discrimination tasks.

During our closed-loop experiment, our purpose was to find the perturbation amplitude with a discrimination of $85\%$. To this end we computed the discrimination probability online as describe above, and then chose the next perturbation amplitude to be displayed using the `accelerated stochastic approximation' method \citep{Kesten58,Faes2007}: when discrimination was above $85\%$, the amplitude was decreased, otherwise, it was increased (see Materials and Methods).

Fig.~2C shows the discrimination probability as a function of the perturbation amplitude for an example perturbation shape. 
Discrimination grows linearly with small perturbations, and then saturates to $100\%$ for large ones. This behavior is well approximated by an error function (gray line) parametrized by a single coefficient, which we call sensitivity coefficient and denote by $c$.
This coefficient measures how fast the discrimination probability increases with perturbation amplitude: the higher the sensitivity coefficient, the easier it is to discriminate responses to small perturbations. 
It can be interpreted as the inverse of the amplitude at which discrimination reaches $76\%$, and is related to the classical sensitivity index $d'$ \citep{macmillan2004}, through $d'=c\times A$, where $A$ denotes the perturbation amplitude (see Materials and Methods). 

All 16 different perturbation shapes were displayed, and the optimal amplitude was searched for each of them independently.
We found a mean sensitivity coefficient of $c=0.0516\ \mu m^{-1}$. However, there were large differences across the different perturbation shapes, with a minimum of $c=0.028\ \mu m^{-1}$ and a maximum of $c=0.065\ \mu m^{-1}$.


\smallskip

\noindent\myparagraph{\bf Sensitivity and Fisher information.}
Can one predict the sensitivity to any perturbation of the stimulus?
The stimulus is the trajectory of a bar and is high dimensional. Generalizing the result of Seung and Sompolinsky \cite{Seung1993} to arbitrary dimension, we show that the sensitivity can be expressed as (see Material and Methods):
\beq\label{eq}
d'=\sqrt{S^{\rm T} \cdot I \cdot S},
\eeq
where $I$ is the Fisher information {\em matrix}, of the same dimension as the stimulus,  and $S$ the perturbation represented as a column vector. Thus, the Fisher information is sufficient to predict the code's sensitivity to any perturbation.

Despite the generality of Eq.~\ref{eq}, it should be noted that estimating the Fisher information matrix requires a model of the population response.
As already discussed, the non-linearities of the retinal code make the construction of a generic model of responses to arbitrary stimuli a very arduous task, and is still an open problem. However, the Fisher information matrix need only be evaluated {\em locally}, around the response to the reference stimulus.

\smallskip

\noindent\myparagraph{\bf Local model for predicting sensitivity.}
We introduce a local model to describe the stochastic population response to small perturbations of the reference stimulus. This model will then be used to estimate the Fisher information matrix, and from it the retina's sensitivity to arbitrary perturbations.

The model, schematized in Fig.~3A, assumes that perturbations are small enough that the response can be linearized around the reference stimulus.
First, the response to the reference is described by conditionally independent neurons firing with time-dependent rates estimated from the peristimulus time histograms (PSTH). 
Second, the response to perturbations is modeled as follows: for each neuron and for each 20 ms time bin of the considered response, 
we use
a linear projection of the perturbation onto a temporal filter
to modify the spike rates relative to the reference.
These temporal filters were inferred from the responses to all the presented perturbations, varying both in shape and amplitude (but small enough to remain within the linear approximation). Details of the model and its inference are given in Materials and Methods.

\begin{figure}
\centering
\includegraphics[width=1\linewidth]{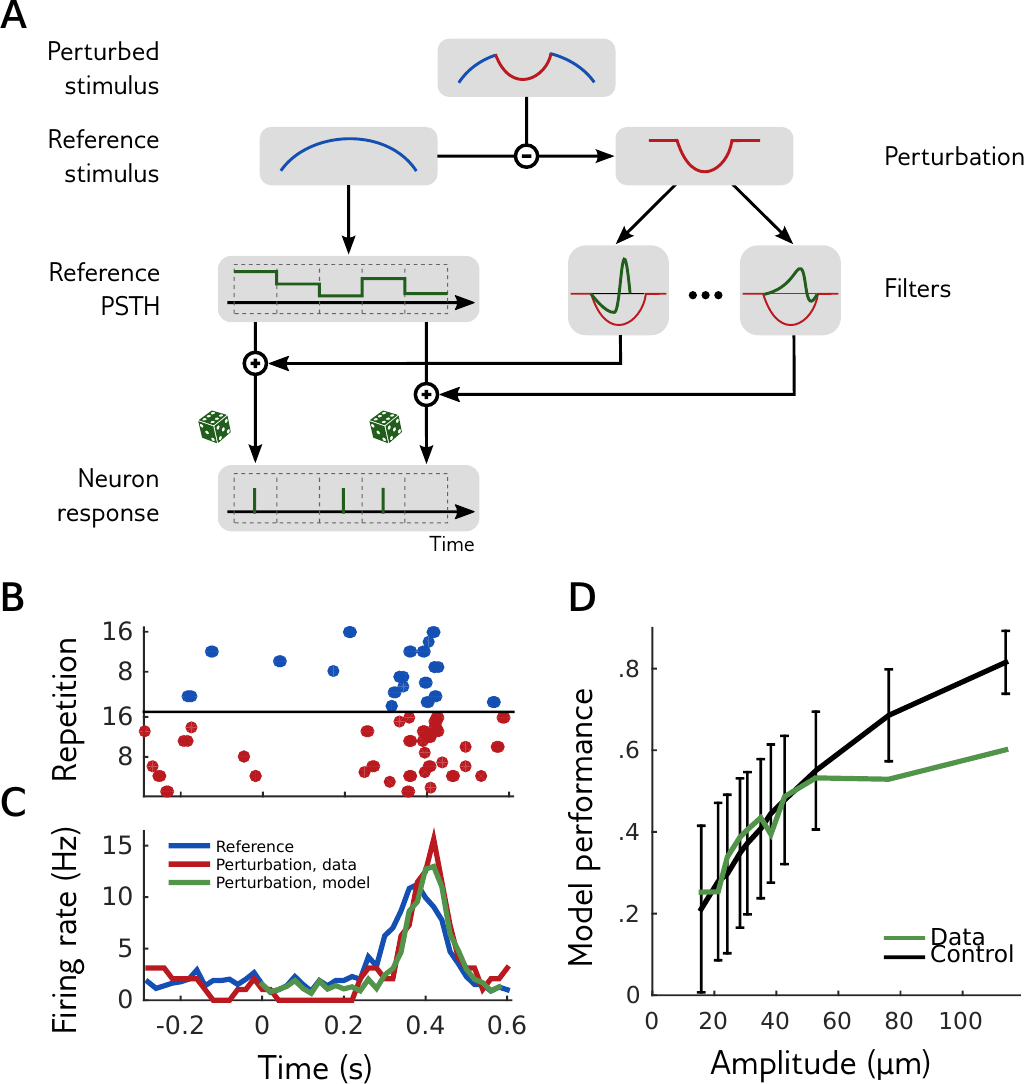}
\caption{
\textbf{Local model for responses to perturbations.}
{\bf A.} The firing rates in response to a perturbation of a reference
stimulus are modulated by filters applied to the perturbation. There
is a different filter for each cell and each time bin.
{\bf B.} Raster plot of the responses of an example cell to the reference (blue) and perturbed (red) stimuli for several repetitions.
{\bf C.} Peristimulus time histogram (PSTH) of the same cell in
response to the same reference (blue) and perturbation (red). Prediction of the local model for the perturbation is shown in green.
{\bf D.} Performance of the local model at predicting the change in
PSTH induced by a perturbation, as measured by Pearson's correlation
coefficient between data and model, averaged over cells (green). The data PSTH were calculated by
grouping perturbations of the same shape and of increasing amplitudes by
groups of 20, and computing the mean firing rate at each time over the 20
perturbations of each group. The model PSTH was calculated by mimicking the same procedure. 
To control for noise from limited sampling, the same performance was
calculated from synthetic data of the same size, where the model is
known to be exact (black).
} 
\label{f:inference}
\end{figure}

We first checked the validity of the local model by its ability to predict the PSTH of cells in response to perturbations (Fig.~3B).
To assess model performance, we computed the difference of PSTH between perturbation and reference, and compared it to the model prediction. 
Fig.~3D shows the correlation coefficient of this PSTH difference between model and data, averaged over all recorded cells for one perturbation shape. To control for noise in the responses, we computed the same quantity for responses generated by the model (black line), which gives an upper bound on the attainable performance given the limited amount of data.
Model performance saturates that bound for amplitudes up to 60 \textmu m, indicating that the local model can accurately predict the statistics of responses to perturbations within that range. For larger amplitudes, the linear approximation breaks down, and the local model fails to accurately predict the response. This failure for large amplitudes is expected if the retinal population responds non-linearly to the stimulus.
We observed the same behavior for all the perturbation shapes tested.

\begin{figure}
\centering
\includegraphics[width=1\linewidth]{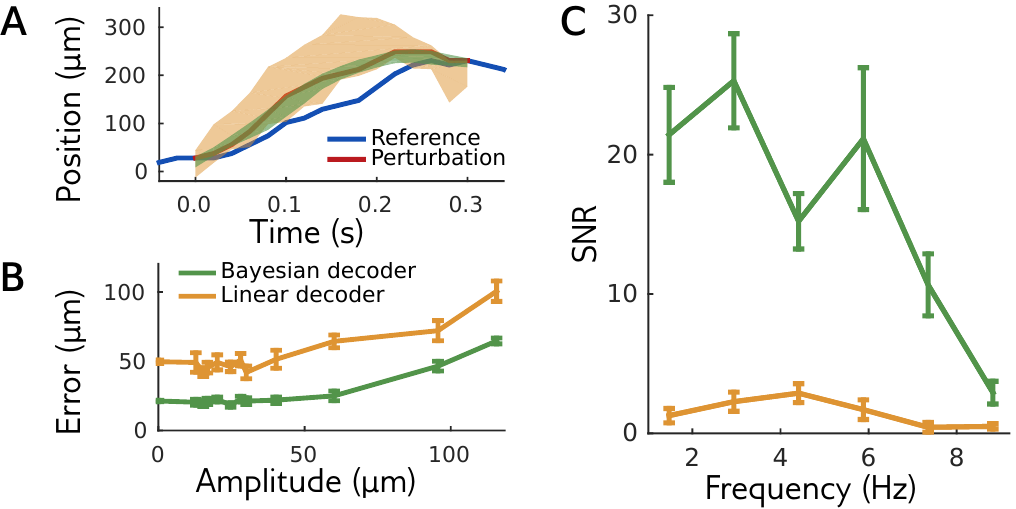}
\caption{
\textbf{Bayesian decoding of the local model outperforms the linear decoder.}
{\bf A.} Responses to a perturbation of the reference stimulus
(reference in blue, perturbation in red) are decoded using the local model (green) or  a linear decoder (orange). For each decoder, the area shows one standard deviation from the mean. 
{\bf B.} Decoding error as a function
of amplitude,  for an example perturbation shape. 
{\bf C.} Signal-to-noise ratio for perturbations with different frequencies (see Materials and Methods). The performance of both decoders decreases for high frequency stimuli.
} 
\label{f:decoding}
\end{figure}

To further test the accuracy of the local model, we estimated its decoding capability, and compared it to the performance of a classical
linear decoder \citep{Warland97,Marre15} trained over an ensemble of random bar trajectories (see Materials and Methods). 
The local model is an encoding model: it predicts the probability of responses given an stimulus. Yet it can be used to create a `Bayesian decoder' using Bayesian inversion (see Materials and Methods): given a response, what is the most likely stimulus that generated this response under the model?
If the local model predicts the retinal response accurately, doing Bayesian inversion of this model should be the best decoding strategy, meaning that other decoders should perform equally or worse. The linear decoder has previously shown high performance in decoding this stimulus \citep{Marre15}, and is therefore a good benchmark for comparison.
When decoding the bar trajectory, the Bayesian decoder was more precise than the linear decoder, as measured by the variance of the reconstructed stimulus
(Fig.~4A). 
The Bayesian decoder had a smaller error than the linear decoder when decoding perturbations of small amplitudes (Fig.~4B).
For larger amplitudes, where the local
model is expected to break down, the performance of the Bayesian decoder decreased.

To quantify decoding performance as a function of the stimulus frequency, we estimated the signal-to-noise ratio (SNR) of the decoding signal for small perturbations of various frequencies (see Materials and Methods). The Bayesian decoder had a much higher SNR than the linear decoder at all frequencies (Fig.~4C; see also Fig.~S3A for an example on another reference stimulus), {even if} both did fairly poorly at high frequencies.
This results suggests that inverting the local model might be the best decoding strategy, and therefore confirms that the local model is an accurate description of the retinal response to small enough perturbations around the reference stimulus.

\begin{figure}
\centering
\includegraphics[width=1\linewidth]{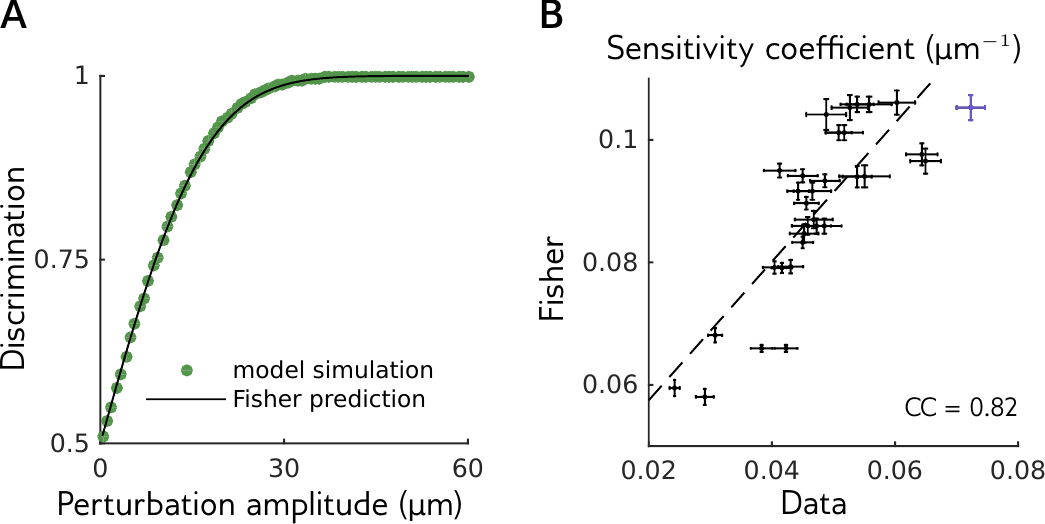}
\caption{
\textbf{The Fisher information predicts the experimentally measured sensitivity.}
{\bf A.} The probability of discrimination of responses to an example perturbation shape
at different amplitudes, calculated from simulations of the local
model (green dots) is well 
predicted by the Fisher information prediction,
$(1/2)[1+\mathrm{erf}(d'/2)]$, with $d'$ given by Eq.~\ref{eq} (black line).
{\bf B.} Sensitivity coefficients $c$ for the two reference stimuli
and 16 perturbation shapes, measured empirically (x axis) and using the
Fisher information (y axis). The purple point corresponds to the perturbation shown in Fig.~2. 
}
\label{f:sensitivity}
\end{figure}

\begin{figure}
\centering
\includegraphics[width=0.95\linewidth]{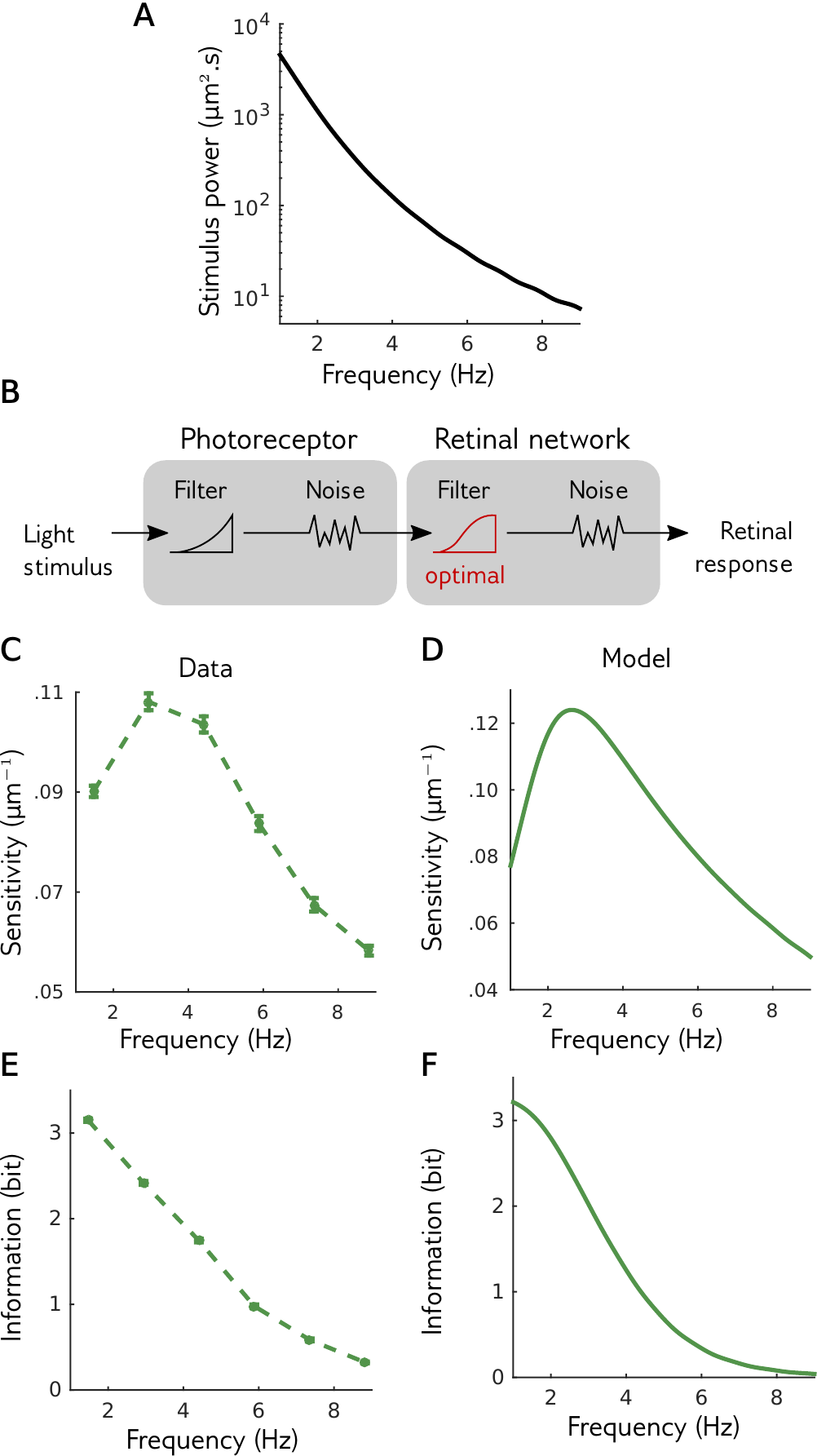}
\caption{
\textbf{Signature of efficient coding in the sensitivity} 
{\bf A.} Spectral density of the stimulus used in experiments, which is monotonically decreasing.
{\bf B.} Simple theory of retinal function: the stimulus is filtered
by noisy photoreceptors, whose signal is then filtered by the
noisy retinal network. The retinal
network filter was optimized to maximize information transfer at
constant output power.
{\bf C.} Sensitivity of the recorded retina to perturbations of different frequencies. Note the non monotonic behavior.
{\bf D.} Same as C, but for the theory of optimal processing.
{\bf E.} Information transmitted by the retina on the perturbations at different amplitudes.
{\bf F.} Same as E, but for the theory.
}
\label{f:efficient}
\end{figure}

\smallskip

\noindent\myparagraph{\bf Ideal versus empirical sensitivity.}
Now that we have validated the local model, we can use it to compute the Fisher information matrix described above.
This matrix takes a very simple form within the local linear model, $I=F\cdot C_R\cdot F^{\rm T}$, where $F$ is the matrix containing the model's temporal filters (stacked as row vectors), and $C_R$ is the covariance matrix of the entire response to the reference stimulus across neurons and time.
This Fisher information is consistent with the sensitivity that we computed earlier. To demonstrate this consistency, we first used the local model to simulate a large number of responses to the different perturbations. When estimating the sensitivity of the response with the method describe above, we obtained identical results than if we estimated it directly from the Fisher information using Eq.~\ref{eq}. This confirms that the Fisher information is a compact and reliable way to estimate the sensitivity of a model.

We then asked if the Fisher information could predict the sensitivity measured above from the recorded responses. We compared the sensitivity coefficients measured experimentally to the prediction by the Fisher information, for all perturbations (Fig.~5B). We found that the empirical sensitivity could be well predicted by the Fisher information (Pearson correlation: $0.82$, $p=10^{-8}$). Although the Fisher prediction always overestimates the observed sensitivity, a large fraction of this difference can be attributed to the limited sampling of the responses, as can be shown on simulated data (Fig.~S2).

\smallskip

\noindent\myparagraph{\bf Signature of efficient coding in the sensitivity.}
{We showed that the Fisher information is a complete description of the sensitivity of the retinal response to perturbations around the reference stimulus. We can now use it to study the sensitivity of the retina to perturbations at different frequencies.} 
Fig.~6A represents 
the power spectrum of the bar motion.
Power is maximum at low frequencies, and quickly decays at large frequencies.
In many classical theories of efficient coding, sensitivity is expected to follow a inverse relationship with the stimulus power \cite{Brunel1998,Wei2016}. However, here we found that the sensitivity is bell shaped, with a peak in frequency around 4Hz (Fig.~6C; see also Fig.~S3B for another reference stimulus).

To interpret this non-monotonic sensitivity, we studied a minimal theory of retinal function (similar to \cite{VanHateren92}) to test how efficient coding would reflect on the sensitivity of the retinal response. In this theory, the stimulus is first passed though a low-pass filter, then corrupted by an input noise. This first stage describes processing by photoreceptors \cite{Ruderman1992}. The photoreceptor output is then transformed by a transfer function and corrupted by a second external noise, which mimics the subsequent stages of retinal processing leading to ganglion cell activity. Here the output is reduced to a single continuous signal (Fig.~6B, see Materials and Methods for mathematical details). Note that this theory is linear: we are not describing the response of the retina to any stimulus, which would be highly non-linear, but rather its linearized response to perturbations around a given stimulus, as in our experimental approach.
To apply the efficient coding hypothesis, we assumed that the photoreceptor filter is fixed, and we maximized the transmitted information, measured by Shannon's mutual information, over the transfer function. We constrained the variance of the output to be constant, corresponding to a metabolic constraint on the firing rate of ganglion cells. In this simple and classical setting, this optimal transfer function, and the corresponding sensitivity, can be calculated analytically. Although the power spectrum of the stimulus and photoreceptor output are monotonically decreasing, and the noise spectrum is flat, we found that the optimal sensitivity of the theory is bell shaped (Fig.~6E), in agreement with our experimental findings (Fig.~6C).

One can intuitively understand this bell-shaped sensitivity.
On one hand, in the small frequency regime, sensitivity is small to balance out and to share of information across frequencies. This result is classic: when the input noise is negligible, the best coding strategy for maximizing information is to whiten the input signal to obtain a flat output spectrum, which is obtained by having the squared sensitivity be inversely proportional to the stimulus power. 
On the other hand, at high frequencies, the input noise is too high for the stimulus to be recovered. Allocating sensitivity and output power to those frequencies is therefore a waste of resources, as it is devoted to amplifying noise, and sensitivity should remain low to maximize information. 
A peak of sensitivity is thus found between the high SNR region, where stimulus dominates noise and whitening is the best strategy, and the low SNR region, where information is lost into the noise and coding resources should be scarce.
A result of this optimization is that the information transferred should monotonically decrease with frequency, {just as the input power spectrum does} (Fig.~6F). 
We tested if this prediction was verified in the data. We estimated similarly the information rate against frequency in our data, and found that it was also decreasing monotonically (Fig.~6D). 
The retinal response has therefore organized its sensitivity across frequencies in a manner that is consistent with an optimization of information transmission across the retinal network.

\section*{Discussion}

We have developed a novel approach to characterize experimentally the sensitivity of a sensory network, and tested if this sensitivity was in agreement with the prediction from efficient coding theory. 
We used closed-loop experiments \cite{Bolinger2012} to measure the sensitivity of the retinal network to small perturbations around a given stimulus. Using a local model that predicts well the neural responses to these perturbations, we could determine the sensitivity of the population, defined by the Fisher information matrix. When expressed in the frequency domain, we found that the Fisher information matrix has a particular structure, with a non-monotonic sensitivity as a function of frequency. We showed that this bell-shaped sensitivity curve corresponds to a signature of efficient coding of the photoreceptor responses by the retinal network. 

Our approach circumvents the need to build a non-linear model that would accurately predict responses to complex stimuli, by restricting ourselves to only a small neighbourhood of the stimulus space. This simplification allowed us to go beyond cases where the retinal function can be summarized by a linear or quasi-linear (e.g. `linear-nonlinear' or LN) model. In the retina, efficient coding theory had led to key predictions about the shape of the receptive fields, explaining their spatial extent \cite{Atick1992,Borghuis2008}, or the details of the overlap between cells of the same type \cite{Doi2012,Karklin2011,Liu2009}. However, when stimulated with complex stimuli like a fine-grained image, or irregular temporal dynamics, the retina exhibits a non-linear behaviour \cite{Gollisch2010}. For this reason, up to now, there was no prediction of the efficient theory for these complex stimuli. Our approach allows us to overcome this barrier and to characterize sensitivity for any arbitrarily complex stimulus ensemble, as long as the experiment duration allows for a reasonable exploration of the perturbation space. 

Another possible use of our method would be to test whether the retina efficiently codes for the statistics of the stimulus by adapting to them specifically, or if it remains an efficient code for natural stimuli, and performs sub-optimally for artificial stimuli. In our case, the temporal dynamics of the bar were close to the kind of dynamics encountered in natural scenes \cite{Eizenman1985,Branson2009}. As a result, it is difficult to tease apart the two hypotheses. Future works could address this question by using stimulus statistics that would be strong departure form natural ones, and test if the retina adapts to code these novel statistics more efficiently, or remains an efficient encoder of natural statistics. 

Recently, more elaborate versions of the efficient coding theory have been suggested, with different constraints on the information maximization. For example, it has been suggested that the retinal network could code efficiently the possible future stimuli, constrained on the information carried about the past stimuli \cite{Palmer2013}. Our approach could be used to test if this predictive information is optimized in cases where the retinal responses are not amenable to a linear or LN model. 

More generally, different versions of the efficient coding theory have been proposed to explain the organisation of several areas of the visual system \cite{Bialek2006,Dan1996,Bell1997,Olshausen1996,Karklin2011} and elsewhere \cite{Smith2006,Machens2001,Chechik2006,Kostal2008}. Our approach is general and could be used in other sensory structures to test the validity of this hypothesis. In our case we found that the local model could be linear in the responses as well as in the stimulus, and still predict the responses to small perturbations. In more complex structures, it is possible that non-linear models in the response are necessary. Yet, the linearity in the stimulus would greatly simplify the problem.
Our approach might therefore be useful in general for sensory systems. 

The experimental characterization of the sensitivity of a neural network could be useful for other purposes than testing the efficient coding theory. First, the local model is supposed to be a very good model for the responses to small perturbations around a given stimulus, and as such, a Bayesian decoder based on this model should perform close to optimal at decoding the stimulus. This can be used to test how close to optimal different decoding methods are. In our case, we found that linear decoding, despite its very good performance, was quite far from the performance of the Bayesian inversion of our local model. This result implies that there should exist non-linear decoding methods that outperform linear decoding \cite{Botella-Soler2016}. Testing the optimality of the decoding method is crucial for brain machine interfaces \cite{Gilja2012}: in this case an optimal decoder is necessary to avoid missing a significant amount of information. Building our local model is a good strategy for benchmarking different decoding methods. 

Finally, the estimation of the sensitivity along several dimensions of the stimulus perturbations allows us to define which changes of the stimulus evoke the strongest change in the sensory network, and which ones should not make a big difference. Similar measures could in principle be performed at the perceptual level, where some pairs of stimuli are perceptually indistinguishable, while others are well discriminated. Comparing the sensitivity of a sensory network to the sensitivity measured at the perceptual level could be a promising way to relate neural activity and perception.


\setcounter{figure}{0}
\makeatletter 
\renewcommand{\thefigure}{S\@arabic\c@figure}
\makeatother

\begin{figure}
\centering
\includegraphics[width=1\linewidth]{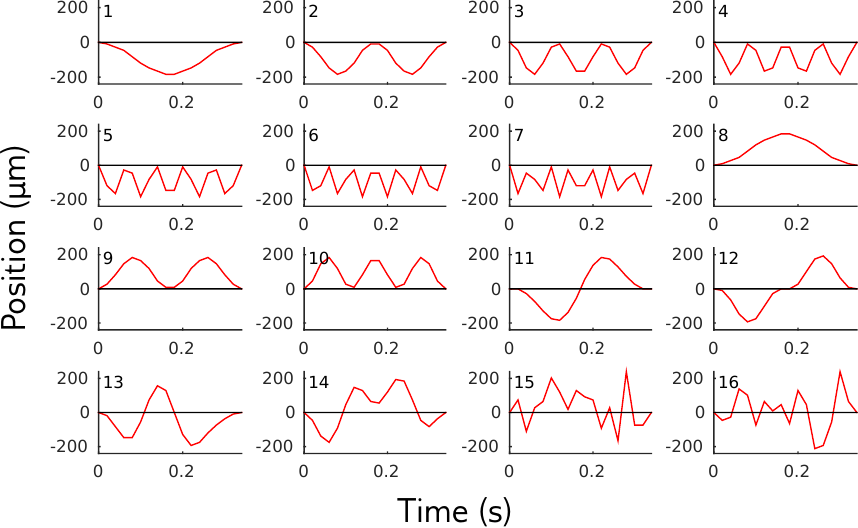}
\caption{
{\bf Perturbations shapes.}
We used the same 16 perturbation shapes for the 2 reference stimuli.
The first 12 perturbation shapes were combinations of simple two Fourier components, and the last 4 ones were random combinations of them:
 $f_k( t )  =  \cos(2 \pi kt /T)$,
$g_k( t )  =  (1/k) \,\sin( 2 \pi kt/T)$,
with $T$ the duration of the perturbation and $t=0$ the beginning of the perturbation. The first perturbations $j=1...7$ were $S_j = f_j  - 1$. For $j = 8,\ldots,10$ they were the opposite of the three first ones: $S_j = - S_{j-7}$. For $j= 11, 12$ we used $S_j = g_{j-10 +1} - g_1$. Perturbations 13 and 14 were random combinations of perturbations 1, 2, 3, 11 and 12, constrained to be orthogonal. Perturbations 15 and 16 were random combinations of $f_j$ for $j \in [1,8]$ and $g_k$ for $k \in [1,7]$, allowing higher frequencies than perturbation directions 13 and 14. Perturbation direction 15 and 16 were also constrained to be orthogonal.
The largest amplitude for each perturbation we presented was 115
\textmu m. An exception was made for perturbations 15 and 16 applied
to the second reference trajectory, as for this amplitude they had a
discrimination probability below 70$\%$. They were thus increased by a
factor 1.5. The largest amplitude for each perturbation was repeated
at least 93 times, with the exception of perturbation 15 (32 times)
and 16 (40 times) on the second reference trajectory.
} 
\label{f:SI_trajectories}
\end{figure}

\begin{figure}
\centering
\includegraphics[width=0.5\linewidth]{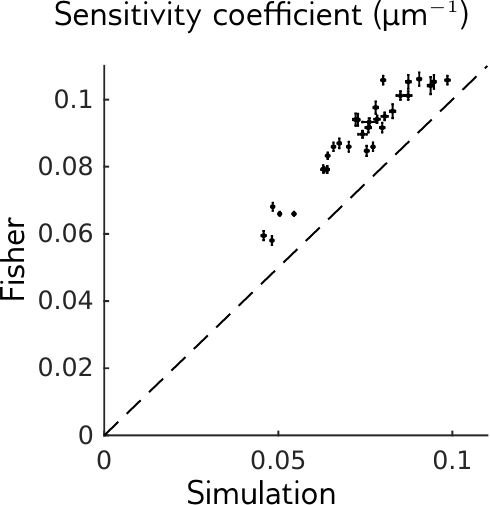}
\caption{
{\bf Simulations with limited data.}
Sensitivity coefficient for different reference stimuli and perturbation shapes, measured with the Fisher information or using simulations of the local model with the same amount of data as in experiments. The discriminability of simulations was measured in the same way than for recorded responses. We show the mean and std over 10 simulation repetitions.
} 
\label{f:SI_simulations}
\end{figure}

\begin{figure}
\centering
\includegraphics[width=.8\linewidth]{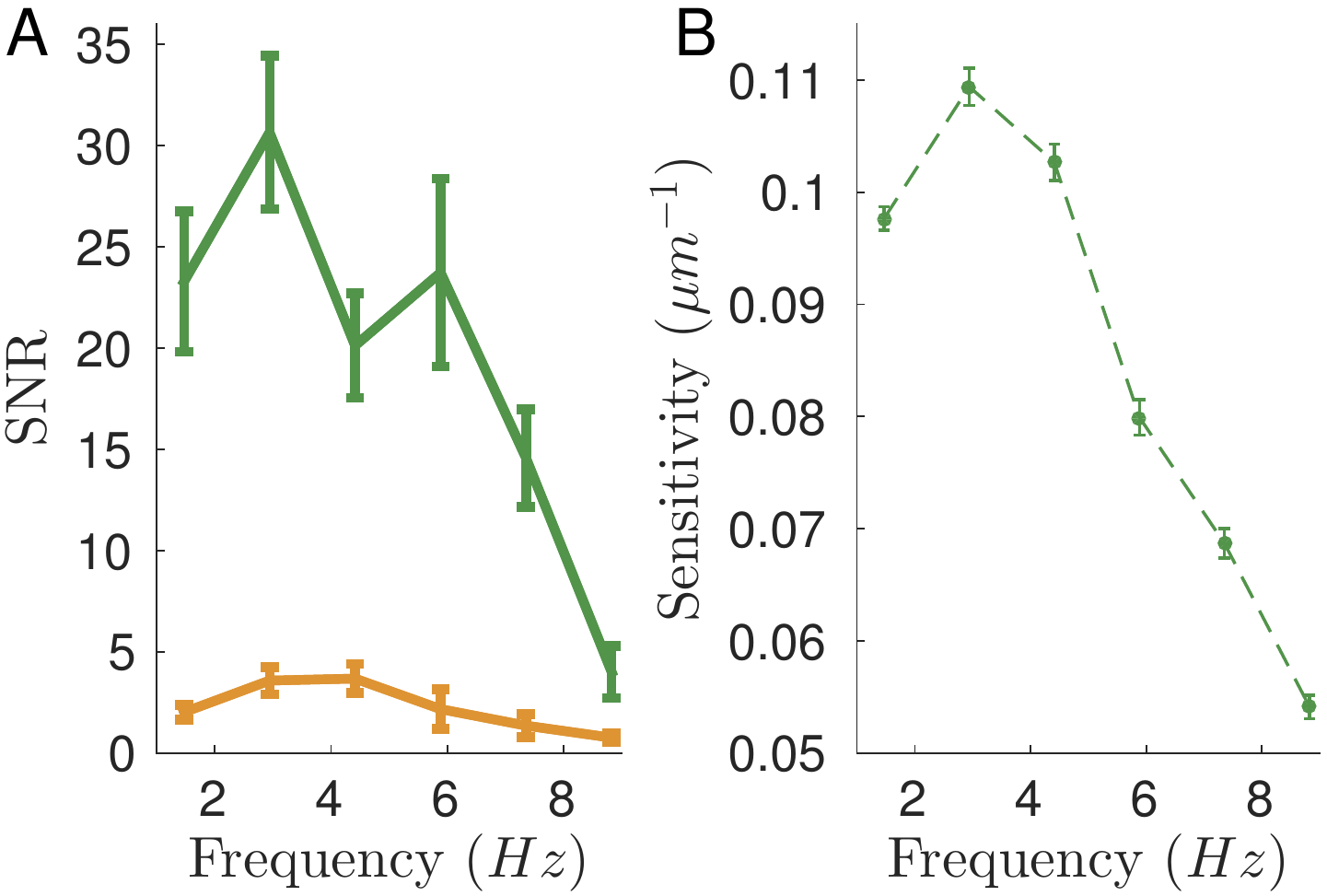}
\caption{
{\bf Results for the second reference stimulus.}
{\bf A.} Same as Fig.~4C of the main text for the second reference
stimulus. {\bf B.} Same as 6C of the main text for the second reference stimulus.
} 
\label{f:SI_simulations}
\end{figure}

\section*{Materials and Methods}
\noindent\paragraph*{\bf Extracellular recording.}
Experiments were performed on the adult Long Evans rat, in accordance with institutional animal care standards. The retina was extracted from the euthanized animal and maintained in an oxygenated Ames' medium (Sigma-Aldrich). 
The retina was recorded extracellularly on the ganglion cell side with an array of 252 electrodes spaced by 60 \textmu m (Multichannel Systems), as previously described \citep{Marre12}.  
Single cells were isolated offline using SpyKING CIRCUS, a {custom} spike sorting algorithm \citep{Yger16}. 
We then selected 60 cells that were well separated (no violations of refractory period), had enough spikes (firing rate larger than 0.5 Hz), had a stable firing rate during the whole experiment, and responded consistently to repetitions of a reference stimulus.  

\noindent\paragraph*{\bf Online spike detection.} 
During the experiment we detected spikes in real time on each electrode independently. Each electrode signal was high-pass filtered using a Butterworth filter with a 200 Hz frequency cutoff. A spike was detected if the electrode potential $U$ was lower than {a threshold of 5 times the median absolute deviation of the voltage}
\citep{Yger16}. 

\noindent\paragraph*{\bf Stimulus.} 
The stimulus was a movie of a white bar on a dark background projected at 50 Hz with a digital micromirror device. The bar had intensity 7.6 10$^{11}$ photons.cm$^{-2}$.s$^{-1}$, and 115 \textmu m width. The bar trajectory consisted in 17034 snippets of 0.9 s consiting in 6431 random trajectories, 2 reference trajectories repeated 391 times each and perturbations of these reference trajectories. All results are shown for the first reference stimulus, except Fig.~5B which combines both. The second reference stimulus served to check the robustness of our results to the choice of reference. Results specific to the second reference stimulus are reported in Fig.~S3.
Continuity between snippets was ensured by constraining all snippets to start and end in the middle of the screen with velocity 0.
Random trajectories followed the statistics of an overdamped stochastic oscillator \citep{Marre15}. We used a Metropolis-Hastings algorithm to generate random trajectories satisfying the boundary conditions. The two reference trajectories were drawn from that ensemble.

A perturbation is denoted by its discretized time series with time step $\delta t=20\ ms$, $S=(S_1,\ldots,S_L)$, with $L=16$, over the $320\ ms$ of the perturbation. Perturbations can be decomposed as $S=A\times P$, where $A^2={(1/L)\sum_{t=1}^L S_t^2}$ is the amplitude, and $P=S/A$ the shape. Perturbations shapes were chosen to have zero value and derivative at their boundaries. They are represented in Fig.~S1.

\noindent\paragraph*{\bf Linear discrimination.}
The response $R$ of the $N=60$ cells over time is binarized into $B$ time bins of size $\delta=20\ ms$: $R_{ib}=1$ if cell $i$ spiked during the $b$th time bin, and 0 otherwise. $R$ is thus a vector of size $N\times B$, labeled by a joint index $ib$. The response is considered from the start of the perturbation until 280 ms afters its end, so that $B=30$.

To discriminate perturbations, we first measured the responses $R_{\rm ref}$ to multiple repetitions of the reference stimulus, and the responses $R_{S_{\rm max}}$ to multiple repetitions of the largest amplitude of each perturbation shape (typically 110 \textmu m).
We computed the mean response to the reference, $\<R_{\rm ref}\>$, and to the largest-amplitude perturbation, $\<R_{S_{\rm max}}\>$, and projected all responses onto their difference: we denote $x_{\rm ref}=(\<R_{S_{\rm max}}\>-\<R_{\rm ref}\>)^{\rm T} \cdot R_{\rm ref}$ the projection of a response to the reference, and $x_{S}=(\<R_{S_{\rm max}}\>-\<R_{\rm ref}\>)^{\rm T} \cdot R_{S}$ the projection of a response to $S$ (when projecting, we recalculated the mean responses by excluding the response to project, to avoid overfitting). The distributions of $x_{\rm ref}$ and $x_S$ are shown in blue and purple in Fig.~2B. The probability of discrimination $D$ is defined as the probability that $x_{\rm ref}<x_S$.

\noindent\paragraph*{\bf Adaptation of perturbation amplitude.}
To identify the range of perturbations that were neither too easy nor too hard to discriminate, we adapted perturbation amplitudes so that the discrimination probability converged to target value $D^*=85\%$. For each shape, perturbation amplitudes were adaptated using the Accelerated Stochastic Approximation \citep{Kesten58}. If an amplitude $A_n$ triggered a response with discrimination probability $D_n$, then at the next step the perturbation was presented at amplitude $A_{n+1}$ with
\beq
\log A_{n+1} = \log A_{n} - \dfrac{ C }{r_n + 1}  ( D_n - D^*  ),
\eeq
where $C=0.74$, and $r_n$ is the number of reversal steps in the experiment, \textit{i.e.} the number of times when a discrimination $D_{n}$ larger than $D^*$ was followed by $D_{n+1}$ smaller than $D^*$, or vice versa. 
We also presented amplitudes regularly spaced on a log-scale. We presented the largest amplitude $A_{max}$ (value in caption of Fig.~S1), and scaled it down by multiples of $1.4$, $A_{max}/1.4^k$ with $k=1,\ldots,7$.

\noindent\paragraph*{\bf Discrimination and sensitivity.}
We assume that the difference of the projections, $\Delta x=x_S-x_{\rm ref}$, is approximately distributed as a Gaussian variable, by virtue of the central limit theorem. For small perturbations, its mean can be assumed to grow linearly with the perturbation amplitude, $\mu=\alpha A$, and its variance $2\sigma^2=\mathrm{Var}(x_S)+\mathrm{Var}(x_{\rm ref})$ to be independent of $A$. Then the probability of discrimination is given by the error function:
\beq\label{discrimination}
D=P( x_{\rm ref}<x_S ) = \dfrac{1}{2} \left(1+ \mathrm{erf} (d'/2) \right)
\eeq
where $d'=\mu/\sigma=c \times A$ is the standard sensitivity index \cite{macmillan2004}, and $c=\alpha/\sigma$ is defined as the sensitivity coefficient, which depends on the perturbation shape $P$. This coefficient determines the amplitude $A=c^{-1}$ at which discrimination is equal to $(1/2)[1+\mathrm{erf}(1/2)]=76\%$.

\noindent\paragraph*{\bf Optimal sensitivity and Fisher information.}
Given the distributions of responses to the reference stimulus, $P(R|{\rm ref})$,  and to a perturbation, $P(R|S)$, optimal discrimination can be achieved by studying the sign of the log-ratio $\mathcal{L}=\ln [P(R|S) / P(R|{\rm ref})]$. Let us call $\mathcal{L}_{\rm ref}$ the value of $\mathcal{L}$ upon presentation of the reference stimulus, and $\mathcal{L}_S$ its value upon presentation of $S$.
The probability of successful discrimination is the probability that $\mathcal{L}_S>\mathcal{L}_{\rm ref}$. Using the central limit theorem we assume again that $\mathcal{L}_S$ and $\mathcal{L}_{\rm ref}$ are Gaussian. We can calculate their mean and variance at small $S$: $\mu=\<\mathcal{L}_S\>-\<\mathcal{L}_{\rm ref}\>=S^{\rm T}\cdot I \cdot S$ and $2\sigma^2=\mathrm{Var}(\mathcal{L}_S)+\mathrm{Var}(\mathcal{L}_{\rm ref}) =2 S^{\rm T}\cdot I \cdot S$, where
\begin{equation}
I_{tt'} = - \sum_{R} P(R|{\rm ref})  \left. \frac{\partial^2 \log P(R |S) }{\partial S_t ~ \partial S_{t'}}\right|_{S=0} \label{eq:fisher}
\end{equation}
is the Fisher information matrix calculated at the reference stimulus.
The discrimination probability is:
$D=P(\mathcal{L}_S>\mathcal{L}_{\rm ref})=(1/2)[1+\mathrm{erf}(d'/2)]$, with 
\beq
d'=\frac{\mu}{\sigma}=\sqrt{S^{\rm T}\cdot I \cdot S}.
\eeq
This result proves \ifthenelse{\boolean{pnas}}{Eq.~1 of the main text}{Eq.~\ref{eq}} with $S=A\times P$.

\noindent\paragraph*{\bf Local model.}
We introduce the local model as a linear expansion of the logarithm of response distribution as a function of both stimulus and response:
\begin{equation}\label{local}
\begin{split}
\log P(R|S) &= \log P(R|{\rm ref}) + \sum_{ib,t} R_{ib}F_{ib, t} S_{t} +{\rm const} \\
&= \log P(R|{\rm ref}) + R^{T}\cdot F\cdot S +{\rm const}.
\end{split}
\end{equation}
The matrix $F$ contains the linear filters with which the change in the response is calculated from the linear projection of the past stimulus. For ease of notation, hereafter we use matrix multiplications rather than explicit sums over $ib$ and $t$.

The distribution of responses to the reference trajectory is assumed to be conditionally independent:
\begin{equation}
\log P(R|{\rm ref}) = \sum_{ib} \log P(\s_{ib} |{\rm ref}).
\end{equation}
Since the variables $R_{ib}$ are binary, their mean value $\<R_{ib}\>$ upon presentation of the reference completely specifies $P(R|{\rm ref})$. They are directly evaluated from the responses to repetitions of the reference stimulus, with a small pseudo-count to avoid zero values.

To infer the filters $F_{ib,t}$, we only include perturbations that are small enough to remain within the linear approximation.
We first separated the dataset into a training ($285 \times 16$ perturbations)  and testing ($20 \times 16$ perturbations) sets. We then defined, for each perturbation shape, a maximum perturbation amplitude above which the linear approximation was considered no longer valid. 
We selected this threshold by optimizing the model's ability to predict the changes in firing rates in the testing set.
Model learning was performed for each cell independently by maximum likelihood with an $L_2$ smoothness regularization on the shape of the filters, using a pseudo-Newton algorithm.
The amplitude threshold obtained from the optimization varied widely across perturbation shapes. The number of perturbations for each shape used in the inference ranged from $20$ ($7\%$ of the total) to $260$ ($91\%$ of the total). Overall only $32\%$ of the perturbations were kept. Overfitting was limited: when tested on perturbations of similar amplitudes, the prediction performance on the testing set was never lower than $15\%$ of the performance on the training set.

Evaluating the Fisher information matrix, Eq. (\ref{eq:fisher}), within the local model, Eq.~\ref{local}, gives:
\beq
I =  F \cdot  C_R \cdot F^{\rm T}
\label{eq:fisherLLM}
\eeq
where $C_R$ is the covariance matrix of $R$, which is diagonal because of conditional independence.

\noindent\paragraph*{\bf Link with linear discrimination.}
There exists a mathematical relation between the Fisher information of Eq.~\ref{eq:fisherLLM} and linear discrimination. The linear discrimination task described earlier can be generalized by projecting the response difference, $R_S - R_{\rm ref}$, along an arbitrary direction $u$:
\beq
\Delta x = x_{S} - x_{\rm ref} =   u^{T} \cdot (R_S - R_{\rm ref}).
\eeq
$\Delta x$ is again assumed to be Gaussian by virtue of the central limit theorem.
We further assume that perturbations $S$ are small, so that $\<R_S\> - \<R_{\rm ref}\>\approx (\partial \<R_S\>/\partial S) \cdot S$, and that $C_R$ does not depend on $S$.
Calculating the mean and variance of $\Delta x$ under these assumption gives an explicit expression of $d'$ in Eq.~\ref{discrimination}:
\beq
d' = \frac{u^{T} \cdot \frac{\partial \<R_S\>}{\partial S}\cdot S}{\sqrt{u^{T}\cdot C_R \cdot u}}.
\eeq
Maximizing this expression of $d'$ over the direction of projection $u$ yields $u= {\rm const}\times C_R^{-1} \cdot (\partial \<R_S\>/\partial S) \cdot S$ and 
\beq
d'=\sqrt{S^{\rm T} \cdot I_L \cdot S},
\eeq
where $I_L=(\partial \<R_S\>/\partial S)^{\rm T}\cdot C_R^{-1}\cdot (\partial \<R_S\>/\partial S)$ is the linear Fisher information \cite{Fisher36,Beck11}. This expression of the sensitivity corresponds to the best possible discrimination based on a linear projection of the response.

Within the local linear model defined above, one has $\partial \<R_S\>/\partial S=F\cdot C_R$, and $I_L=F\cdot C_R\cdot  F^{T}$, which is also equal to the true Fisher information (Eq.~\ref{eq:fisherLLM}): $I=I_L$. Thus, if the local model (Eq.~\ref{local}) is correct, discrimination by linear projection of the response is optimal and saturates the bound given by the Fisher information.

Note that the optimal direction of projection only differs from the direction we used in the experiments, $u=\<R_S\>-\<R_{\rm ref}\>$,  by an equalization factor $C_R^{-1}$. We have checked that applying that factor only improves discrimination by a few percents (data not shown).

\noindent\paragraph*{\bf Linear decoder.}
We infer the linear decoder filters  by minimizing the mean square error \citep{Warland97} in the reconstruction of $4000$ random trajectories governed by the dynamics of an overdamped oscillator with noise.
To allow for a meaningful comparison with the local model, the linear filters integrate the past of the stimulus for $\tau=15\times\delta t = 300\, ms$.
Tested on a sequence of $\sim 400$ repetitions of one of the two reference trajectories, where the first $300\  ms$ of each have been cut out, we obtain a correlation coefficient of $0.87$ among the stimulus and its reconstruction.

\noindent\paragraph*{\bf Bayesian decoder.}
We use Bayes' rule to infer the presented stimulus given the response:
\begin{equation}
P(S|R) = \frac{P(R|S)P(S)}{P(R)}  \label{posterLLM}
\end{equation}
where $P(R|S)$ is given by the local model (Eq.~\ref{local}), $P(S)$ is the prior distribution over the stimulus, and
$P(R)$ is a normalization factor that does not depend on the stimulus. $P(S)$
is taken to be the distribution of trajectories from an overdamped stochastic oscillator with the same parameters as in the experiment.
The stimulus is inferred by maximizing the posterior $P(S|R)$ numerically, 
using a pseudo-Newton iterative algorithm.

\noindent\paragraph*{\bf Local signal to noise ratio in decoding.}
For any decoder, the inferred value of the stimulus $\hat S$ can be written as:
\begin{equation}
\hat S = T \cdot S + b + \epsilon,
\label{eq:DecodingEffective}
\end{equation}
where $T$ is a transfer matrix which differs from the identity matrix when decoding is imperfect, $b$ is a systematic bias,
$\epsilon$ is a Gaussian noise of covariance $C_\epsilon$.
We inferred the values of $b$ and $C_\epsilon$ from the $\sim 400$ reconstructions of the reference stimulation using either of the two decoders, and the values of $T$ from the reconstructions of the perturbed trajectories.
The inference is done by an iterative algorithm similar to that used for the inference of the filters $F$ of the local model.
The signal-to-noise ratio (SNR) in decoding the perturbation  $S$ is then defined as:
\begin{equation}
\textrm{SNR}(S) = S^{\rm T} \cdot T^{\rm T} \cdot C_\epsilon^{-1} \cdot T \cdot  S.
\label{eq:SNR}
\end{equation}
In Fig.~4C, to compute SNR for a frequency $\nu$, we use Eq.~\ref{eq:SNR} with
$S_t = A~  \exp( 2 \pi i \nu t \delta t)$,
where $A$ is the amplitude of the perturbation shown in Fig.~4A.

\noindent\paragraph*{\bf Frequency dependence of sensitivity and information.}
To analyze the behavior in frequency of the sensitivity, we compute the sensitivity index for an oscillating perturbation of unitary amplitude.
We apply \ifthenelse{\boolean{pnas}}{Eq.~1 of the main text}{Eq.~\ref{eq}} with $\hat S_t(\nu) \equiv \exp( 2 \pi i \nu t \delta t)$.
In order to estimate the spectrum of the information rate we compute its behavior within the linear theory \cite{VanHateren92}:
\begin{equation}
\text{MI}(\nu) = \frac{1}{2} \log\left[ 1 + C_S(\nu) I(\nu) / \delta t^2 \right]
\end{equation}
where $C_S(\nu)$ is the power spectrum of stimulus, and $I(\nu)=(\delta t / L)\hat S^{\rm T}(\nu)\cdot I\cdot \hat S(\nu)$.

\noindent\paragraph*{\bf Efficient coding theory.}
To build a theory of retinal sensitivity, we follow closely the approach of Van Hateren \cite{VanHateren92}.
The stimulus is first  linearly convolved with a filter $f$, of power $\mathcal{F}$, then corrupted by an input white noise with uniform power $H$, then convolved with the linear filter $r$ of the retina network of power $\mathcal{G}$, and finally corrupted again by an external white noise $\Gamma$.
The output power spectrum $O(\nu)$ can be expressed as a function of frequency $\nu$:
\begin{equation}
O(\nu) = (\delta t L)\mathcal{G}(\nu)[(\delta t L) \mathcal{F}(\nu) C_S(\nu) + {H} ]  + \Gamma
\end{equation}
where $C_S(\nu)$ is the power spectrum of the input.
The information capacity of such a noisy input-output channel is limited by the allowed total output power $V=\sum_\nu O(\nu)$, which can be interpreted as a constraint on the metabolic cost.
The efficient coding hypothesis consists in finding the input-output relationship $g^*$, of power  $\mathcal{G}^*(\nu)$, that maximizes the information transmission under a constraint on the total power of the output.
The optimal Fisher information matrix can be computed in the frequency domain as:
\begin{equation}
I(\nu)= \frac{\delta t^4 L^2 \mathcal{G}^*(\nu) \mathcal{F}(\nu)}{\Gamma + L \delta t \mathcal{G}^*(\nu) {H}}.
\end{equation}
The photoreceptor filter \cite{Warland1997} was taken to be exponentially decaying in time, $f = \tau^{-1} \exp(-t/\tau)$ (for $t\geq 0$), with $\tau = 100\ ms$.
The curve $I(\nu)$ only depends on $H$, $\Gamma$ and $V$ through two independent parameters.
For the plots in \ifthenelse{\boolean{pnas}}{Fig.~6 of the main text}{Fig.~\ref{f:efficient}} we chose: ${H}=3.38\ \mu m^2 s$, $\Gamma=0.02\  \text{spikes}^2 s$ and $V= 307\ \text{spikes}^2 s$, $\delta t=20\ ms$, and $L=2,500$.
In \ifthenelse{\boolean{pnas}}{Fig.~6}{Fig.~\ref{f:efficient}}D, we plot the sensitivity to oscillating perturbation with fixed frequency $\nu$, which results in $\sqrt{I(\nu) L/\delta t}$.
In \ifthenelse{\boolean{pnas}}{Fig.~6}{Fig.~\ref{f:efficient}}E we plot the spectral density of the transferred information rate:
\begin{equation}
\text{MI}(\nu) = \frac{1}{2} \log\left[ 1 + \frac{ (\delta t L)^2\mathcal{G}(\nu) \mathcal{F}(\nu) C_S(\nu) }{\Gamma + (\delta t L) \mathcal{G}(\nu){H} }\right].
\end{equation}

\bigskip
~

\bigskip

{\bf Ackownledgements.} 
We thank St\'ephane Deny for his help with the
experiments, and Jean-Pierre Nadal for stimulating discussions and
crucial suggestions.
This work was supported by ANR TRAJECTORY, ANR
OPTIMA, the French State program Investissements d'Avenir managed by
the Agence Nationale de la Recherche [LIFESENSES: ANR-10-LABX-65],
European Comission grant from the Human Brain Project n. FP7-604102,
and National Institutes of Health grant n. U01NS090501.

\bibliographystyle{pnas}

\end{document}